\documentclass[aip, reprint]{revtex4-1}

\usepackage{import}
\usepackage{amsmath,amssymb,amsfonts}
\usepackage{graphicx}
\usepackage{hyperref}

\begin{document}


\title{
The Impact of Feature Representation \\
on the Accuracy of Photonic Neural Networks
}

\author{Mauricio Gomes de Queiroz}
\email[Author to whom correspondence should be addressed: ]{mauricio.gomes@ec-lyon.fr}
\affiliation{Ecole Centrale de Lyon, INSA Lyon, CNRS, Universite Claude Bernard Lyon 1, CPE Lyon, INL, UMR5270, 69130 Ecully, France}

\author{Paul Jimenez}
\affiliation{Ecole Centrale de Lyon, INSA Lyon, CNRS, Universite Claude Bernard Lyon 1, CPE Lyon, INL, UMR5270, 69130 Ecully, France}

\author{Raphael Cardoso}
\affiliation{Ecole Centrale de Lyon, INSA Lyon, CNRS, Universite Claude Bernard Lyon 1, CPE Lyon, INL, UMR5270, 69130 Ecully, France}

\author{Mateus Vidaletti Costa}
\affiliation{Ecole Centrale de Lyon, INSA Lyon, CNRS, Universite Claude Bernard Lyon 1, CPE Lyon, INL, UMR5270, 69130 Ecully, France}
\affiliation{School of Engineering, RMIT University, Melbourne, VIC 3000, Australia}

\author{Mohab Abdalla}
\affiliation{Ecole Centrale de Lyon, INSA Lyon, CNRS, Universite Claude Bernard Lyon 1, CPE Lyon, INL, UMR5270, 69130 Ecully, France}
\affiliation{School of Engineering, RMIT University, Melbourne, VIC 3000, Australia}

\author{Ian O'Connor}
\affiliation{Ecole Centrale de Lyon, INSA Lyon, CNRS, Universite Claude Bernard Lyon 1, CPE Lyon, INL, UMR5270, 69130 Ecully, France}

\author{Alberto Bosio}
\affiliation{Ecole Centrale de Lyon, INSA Lyon, CNRS, Universite Claude Bernard Lyon 1, CPE Lyon, INL, UMR5270, 69130 Ecully, France}

\author{Fabio Pavanello}
\affiliation{Univ. Grenoble Alpes, Univ. Savoie Mont Blanc, CNRS, Grenoble INP, CROMA, 38000, Grenoble, France}

\date{\today}

\begin{abstract}

Photonic Neural Networks (PNNs) are gaining significant interest in the research community due to their potential for high parallelization, low latency, and energy efficiency. PNNs compute using light, which leads to several differences in implementation when compared to electronics, such as the need to represent input features in the photonic domain before feeding them into the network. In this encoding process, it is common to combine multiple features into a single input to reduce the number of inputs and associated devices, leading to smaller and more energy-efficient PNNs. Although this alters the network's handling of input data, its impact on PNNs remains understudied. This paper addresses this open question, investigating the effect of commonly used encoding strategies that combine features on the performance and learning capabilities of PNNs. Here, using the concept of feature importance, we develop a mathematical methodology for analyzing feature combination. Through this methodology, we demonstrate that encoding multiple features together in a single input determines their relative importance, thus limiting the network's ability to learn from the data. Given some prior knowledge of the data, however, this can also be leveraged for higher accuracy. By selecting an optimal encoding method, we achieve up to a 12.3\% improvement in accuracy of PNNs trained on the Iris dataset compared to other encoding techniques, surpassing the performance of networks where features are not combined. These findings highlight the importance of carefully choosing the encoding to the accuracy and decision-making strategies of PNNs, particularly in size or power constrained applications.

\end{abstract}

\maketitle 

\section{Introduction}
\label{Introduction}
    \begin{figure*}[t]
    \centering
    \includegraphics[width=\linewidth]{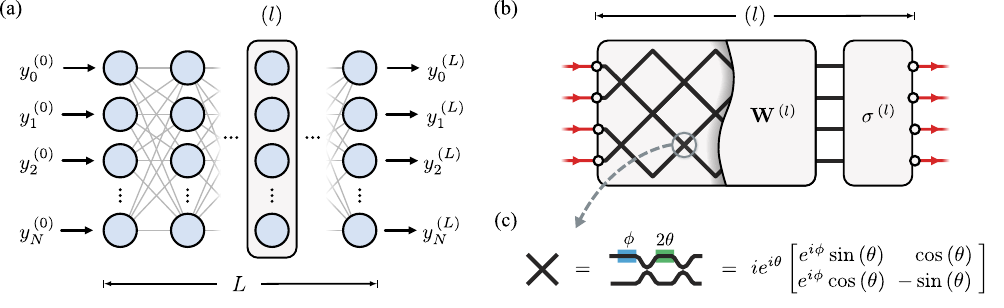}
    \caption{(a) Representation of a generic neural network. An arbitrary layer $l$ is highlighted. (b) Schematic of the photonic implementation of a neural network layer using meshes of Mach-Zehnder Interferometers (MZIs). (c) An MZI and its two phase shifters $(\phi,2\theta)$ are illustrated  alongside the transfer matrix representation of the transformation it performs over the field amplitude.}
    \label{fig:layerANNandPNN}
\end{figure*}

Artificial Intelligence (AI) systems gained widespread relevance in recent years\cite{dong2021survey}, finding diverse applications ranging from image classification \cite{simonyan2014very} to speech recognition \cite{purwins2019deep}. 
These systems have traditionally been implemented on electronic hardware, benefiting from the steady performance improvements driven by the miniaturization of electronic integrated circuits.
However, with components now shrinking to the atomic scale, the limitations of this platform become apparent \cite{theis2017end}.
At this size, for example, quantum effects may disrupt functionality \cite{powell2008quantum}, and the heat from densely packed devices becomes hard to dissipate \cite{leiserson2020room}.
In response, new technologies are being explored to enable further improvements in AI. These emerging technologies are often not subject to the same constraints of their electronic counterparts, and thus might offer more efficient alternatives for certain applications \cite{waldrop2016chips}.

Photonic Neural Networks (PNNs) are hardware implementations of AI systems that perform computations on optical signals, rather than on electronic ones. Using light, they are able to leverage several of its properties to potentially enable high parallelization, low latency, and reduced power consumption \cite{shen2017deep}. For example, PNNs have been demonstrated to perform sub-nanosecond image classification \cite{ashtiani2022chip} and to achieve up to $10^{12}$ Multiply-Accumulate operations per second \cite{xu202111}.
However, transitioning from electronics to photonics remains challenging. Practical applications of medium to large-scale systems are currently limited by the large physical footprint of photonic circuits \cite{shibata2008compact, xiao2021large}, their loss accumulation, and the high power consumption of some of its electro-optic devices \cite{tait2022quantifying}.

One way of alleviating these issues is by optimizing circuits \cite{mourgias2022noise,de2023power}, or carefully designing PNNs to minimize circuit size.
A common practice found in literature involves taking advantage of the complex representation of light (using amplitude and phase) to represent multiple features in a single input, thus combining multiple real-valued features into fewer complex-valued inputs.
By using fewer inputs, a circuit requires fewer components and a smaller network, which leads to a reduction in overall footprint. Such technique aligns well with the capabilities of photonic circuits, which are able to process complex inputs through complex transformations \cite{zhang2021optical}.

However, the way we represent features in Neural Networks (NNs) greatly influences the difficulty of the problems they solve. For example, in tasks with radial symmetry centered around the origin, opting for a polar coordinate system can emphasize the relevant feature relationships necessary for accurately solving the task, short-cutting the network's need to learn it. This approach can significantly reduce the computational complexity required for achieving high accuracy.
Moreover, the choice of feature representation also shapes the network's approaches to solve tasks, as NNs tend to rely on the most straightforward cues available within the data \cite{geirhos2020shortcut}. 
This highlights the need to understand which feature relationships are emphasized by the representation strategies used in PNNs. By doing so, we can ensure that these networks not only achieve high accuracy, but also adopt desirable decision-making strategies.

In this paper, we explore the role of feature representation in the accuracy and decision-making strategies of PNNs. We investigate the common practice of combining various features into a single input, using eXplainable AI (XAI) methods to compare relative importances of the combined features. To our knowledge, only one work investigated such practice as a means of improving accuracy in PNNs \cite{qiu2024oplixnet}. However, the consequences of the feature combination itself are still unknown, and different feature representations were not explored. 
Our work tackles these open questions with a mathematical analysis of feature combination focused on photonic implementations, where networks and circuits are constrained by size. We point out how different data representations and hardware implementations can be exploited for higher accuracy and lower complexity, as well as the shortcomings of current solutions.

The rest of this paper is structured as follows: in Sections \ref{Photonic Neural Networks} and \ref{Feature Importance} we review the basics of photonic implementations of AI and feature importance metrics. In Section \ref{Methodology}, we calculate the relative importance of features that share a same input.
Sections \ref{n-sphere} and \ref{Simulations} discuss practical examples and simulations of Artificial Neural Networks (ANNs) and PNNs. Finally, Section \ref{Conclusions} concludes the discussions brought up in this paper.

\section{Photonic Neural Networks}
\label{Photonic Neural Networks}
    In this section, we provide a review of ANNs and their photonic implementations.
We also address common strategies of representing features in light, which will be used in our further discussions in Section \ref{Methodology}. 

\subsection{Artificial Neural Networks}

Artificial Neural Networks (ANNs), first proposed in the 1940s \cite{mcculloch1943logical}, are mathematical functions loosely inspired by how the human brain processes information. These functions are known to be universal approximators \cite{hornik1989multilayer}, hence their ability to handle a wide variety of tasks.
The network’s behavior, i.e. the way it processes inputs, is determined by their connection strengths (called ``weights'') and  non-linearities (referred to as ``activation functions'') \cite{sze2017efficient}. Typically, these parameters are obtained through training, approximating the ANN to a probability function associated with the  given task.
For instance, in classification tasks, ANNs are designed to assign a class to an input, by approximating a function that calculates the likelihood of belonging to each class \cite{goodfellow2016deep}.

Consider a fully connected, feed forward (FF) NN, consisting of $L$ layers and designed with $N$ inputs and $N$ outputs, as depicted in Fig.~\ref{fig:layerANNandPNN}(a). The process by which a given layer $l$ transforms its inputs is described as:
\begin{align}
\vec{z}^{\, (l)} &= \mathbf{W}^{\, (l)} \cdot \vec{y}^{\, (l-1)} + \vec{b}^{\, (l)} \, ,\\
\vec{y}^{\, (l)} &= \sigma^{\, (l)}(\vec{z}^{\, (l)}) \, .
\end{align}

Initially, inputs are combined through weighted sums by a weight matrix $\mathbf{W}^{\, (l)}$ to obtain $z^{\, (l)}$. Then, the element-wise application of an activation function $\sigma(\cdot)$ to $z^{\, (l)}$ introduces non-linearity and yields the output of the layer, where ${y}^{\,(0)}$ is the input of the network and ${y}^{\,(L)}$ the output. A bias $\vec{b}$ might be added before the activation function to allow for the network to better adjust to the data. 
The entire network, from first to last layer, can be seen as a sequence of such transformations, written as $\vec{y}^{\,(L)} = f(\vec{y}^{\,(0)})$.

Thus, ANNs implement input-output mappings that can be either real or complex. Real-Valued Neural Networks (RVNNs) are characterized by real parameters and inputs, with $f:\mathbb{R}^{N} \mapsto \mathbb{R}^{N}$. In these networks, each layer scales and combines inputs before non-linearly transforming them. 
Complex-Valued Neural Networks (CVNNs), on the other hand, operate in the complex domain, meaning that both the input vector and the network's parameters are complex-valued and $f : \mathbb{C}^{N} \mapsto \mathbb{C}^{N}\,$ \cite{bassey2021survey}. In that case, each layer has the ability to not only scale and combine, but also rotate inputs in the complex plane. This rotation, inherent to complex algebra, makes CVNNs more suitable for tasks where phase information is important, such as in audio processing \cite{kim2024sound} or optical communications \cite{masaad2023photonic}.

\subsection{Photonic Implementations}

Photonic computing is emerging as a promising approach to improve ANN implementations for specific applications by computing with light.
This allows us to leverage its unique characteristics to potentially enable faster and more energy-efficient AI systems. For example, in the optical domain, linear transformations can be done passively \cite{Reck1994} and information can be easily parallelized and processed at high speeds \cite{xu202111}.

PNNs are implementations of ANNs through photonic inputs, components, and transformations \cite{shastri2021photonics}. Although no single photonic component acts as an artificial neuron, a circuit can be designed to perform the mathematical operations of an ANN. This is achieved by using several components such as waveguides, interferometers, and modulators which guide and manipulate light signals. These circuits operate on complex signals and implement complex transformations, meaning that PNNs can act as RVNNs and CVNNs, depending on the task at hand.

Several PNN circuits were suggested and demonstrated experimentally. They can be broadly categorized by how different inputs are distinguished, whether through spatial, wavelength, or time domains \cite{bai2023photonic}.

In this study, we focus on PNNs that use spatial differentiation of inputs. These networks assign a separate input to each optical signal, and implement weight matrix multiplications by making different inputs interfere with each other. Most notably, this is achieved by using meshes of Mach-Zehnder Interferometers (MZIs) \cite{Reck1994, Clements2016}. The interference, and hence the specific mathematical operation performed by the mesh, can be selected by adjusting the phase shifters found in these devices. 
Activation functions, on the other hand, can be implemented by using any of the devices and circuits that exhibit optical non-linearity \cite{ian2020reprogrammable, aashu2020reconfigurable}. The schematics of an ANN implementation and an MZI are shown in Fig.~\ref{fig:layerANNandPNN}(b) and Fig.~\ref{fig:layerANNandPNN}(c), respectively.

If PNNs use coherent light inputs, they can be represented in the complex domain. In these networks, the $i^{\text{th}}$ input is characterized by an amplitude $A_i$ and phase $\phi_i$. Thus, the input vector can be expressed as $\vec{y}^{\,(0)} = \left[ A_1e^{i\phi_1}, \cdots, A_N e^{i\phi_N} \right]^{\intercal} \in \mathbb{C}^{N}$.
Given the two degrees of freedom available for each input, feature encoding can be achieved using various methodologies. We divide common approaches found in literature into two distinct groups: real and complex encoding.

Real encoding simplifies the input representation by encoding data solely in the amplitude of the optical signals, maintaining a uniform initial phase across all inputs (in practice having $\phi_i = 0\,\forall\,i$ and thus $\vec{y}^{\,(0)} \in \mathbb{R}^{N}$). Several researchers employ this encoding method for its compatibility with RVNNs used in electronic computers \cite{shen2017deep,mojaver2023addressing}. It allows for an easy mapping of weights from electronically trained networks to photonic transformations. In these networks, while the nature of the transformations of individual MZIs is inherently complex, the overall behaviour can effectively be real-valued. Since no phase information is used, only the amplitude of the outputs is of interest, which simplifies the detection scheme. However, it is important to ensure that different inputs experience the same phase before reaching the network to maintain phase consistency, which might not be simple to achieve experimentally.  

In contrast, complex encoding uses both amplitude and phase at the same time, having inputs that lie in the complex plane, that is $\vec{y}^{\,(0)} \in \mathbb{C}^{N}$. The transformations in the PNN in this case are complex, and thus, detection of both intensity and phase in the outputs might be used, adding to the electronic complexity of the circuit. In image classification tasks, for example, real-valued input images can be transformed into Fourier space representation to obtain phase and amplitude information \cite{Banerjee2023, hamerly2022asymptotically,wang2022Multicore}, or have different sections mapped to the real and imaginary parts of complex numbers \cite{Fang2019, qiu2024oplixnet}, which reduces by half the number of inputs. 

The encoding choice for PNNs influences the network's behaviour, the type of information that is detected at the output, and the overall size of the circuit, as it may imply the use of additional peripheral devices.
Beyond hardware specifications, this choice might also impact how features are processed within the network. When two features share the same input, the network may process them differently from the way they would be processed individually.
Understanding these dynamics is crucial for optimizing PNN performance.

\section{Feature Importance}
\label{Feature Importance}
    In this section, we look to the field of XAI for methods of evaluating feature importance in ANNs, to later study the impact of combining features in PNNs. We focus on gradient-based techniques, particularly sensitivity analysis.

ANNs, especially those with several layers, are highly non-linear models that use numerous parameters. The network's complexity often leads them to be regarded as opaque or ``black-box" systems, since their decision-making processes are difficult to grasp intuitively. That is, while we can mathematically describe how a given output is obtained, it is difficult to specify ``why" with an intuitive explanation.

Nonetheless, being able to explain the decision-making strategies of a model has a number of practical applications.
Clear explanations can, for instance, enhance our understanding of a problem or be used to demonstrate fair treatment.
In photonics research, XAI is currently used to explain the inverse design of circuits \cite{zhetao2023inverse}, or to aid in the description of physical models \cite{yeung2020elucidating}.
The concept of ``explainability" is still subject of an ongoing debate \cite{lipton2018mythos, miller2019explanation} and, consequently, a variety of methods have been proposed to attain it \cite{samek2021explaining, montavon2018methods}.
Highlighting which input features are considered as important to an ANN is a common way to explain its outputs. Several methods estimate such feature importance, of which we emphasize sensitivity analysis.

Sensitivity analysis quantifies feature importance by examining how sensitive the output of the model is to small variations in each feature \cite{sung1998ranking, fu1993sensitivity, dimopoulos1999neural}. 
The underlying principle is that if small changes in an input lead to significant changes in the output, then that input is likely to be important for the network, i.e. it contributes to the prediction of this output.
In such case, the importance of the $i^{\text{th}}$ input, $y^{\, (0)}_i$, to the the $c^{\text{th}}$ output of the network, $y^{\, (L)}_c$, is denoted by $ R_{i \to c}$:
\begin{equation}\label{eq:sensitivity}
    R_{i \to c} = 
    \left| \frac{\partial y^{\, (L)}_c }{\partial y^{\, (0)}_i} \right| \, .
\end{equation}

Gradient-based explanations are frequently used in image classification tasks to generate saliency maps \cite{Simonyan14a}, and also show fair performance in matching feature importance in simulated data \cite{olden2004accurate}.
Over time, other methods built up on sensitivity analysis, addressing some of its drawbacks  by suggesting additional forms of estimating feature importance \cite{ancona2017towards}. For example, adding Gaussian noise to the input and averaging their resulting gradients helps generating more consistent saliency maps \cite{smilkov2017smoothgrad}.
These techniques are often easy to implement, given that the necessary partial derivatives can be computed through back-propagation. 

\section{Analytical Derivation of Relative Importance}
\label{Methodology}
    Here, we use the concepts elaborated in previous sections to investigate how the importance of features is shaped in PNNs. Initially, we employ the sensitivity analysis shown in Section \ref{Feature Importance} to obtain the importance of an arbitrary feature encoded in one input. Then, we introduce the concept of \emph{encoding functions} to describe the different feature encoding processes and representations in photonics, seen in Section \ref{Photonic Neural Networks}.

Consider a set of features $\mathcal{X} = \{x_1, \cdots, x_n\} \in \mathbb{R} $.
Assume that we want all of the elements in $\mathcal{X}$ to be used by our model. However, due to either a prohibitively large quantity of features or size restrictions on our network, we also wish to use a number of inputs that is less than the number of elements in $\mathcal{X}$.
To achieve both objectives, we combine features into complex inputs, as seen in Section \ref{Photonic Neural Networks}. In this context, we calculate the relative importance of such combined features, to understand what relationships are highlighted by our inputs. 

\begin{figure*}[t]
    \centering
    \includegraphics[width=\linewidth]{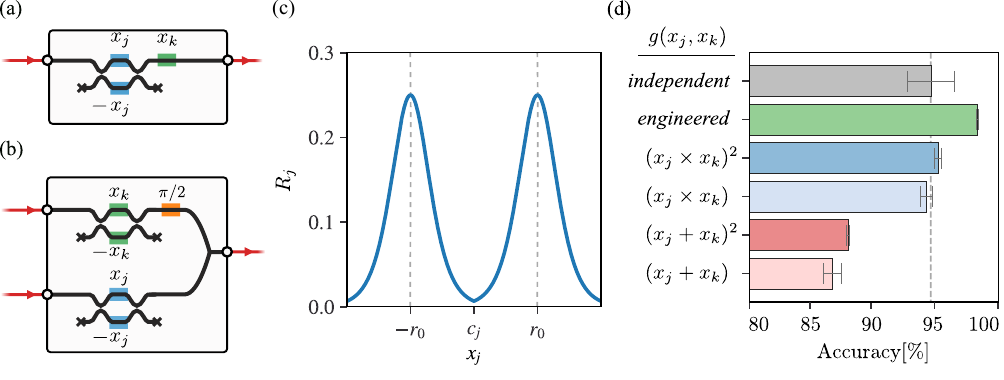}
    \caption{(a) and (b) show photonic circuits that implement encoding functions. By modulating the phase shifters to the indicated value, one can achieve functions similar to exponential encoding (a) or linear encoding (b). (c) Importance of a feature $x_j$ when $r_p = |x_j|$, which is equal to moving $P$ alongside the $x_j$ axis. (d) Mean accuracy of $100$ ANNs trained with several encoding functions. ``Independent" refers to the situation where features are not combined, and ``engineered" to the implementation of Eq.~\eqref{eq:nsphere:real_representation}.}
    \label{fig:nsphere:big}
\end{figure*}  

Our first objective is to obtain the importance of an arbitrary feature, say $x_j \in \mathcal{X}$, with respect to an arbitrary output of the model. This feature is represented only in input $y^{\, (0)}_i$, and its importance is assessed in relation to the $c^{\text{th}}$ output of the network. Applying the chain rule to Eq.~\eqref{eq:sensitivity}, we write this importance $R_{j \to c}$ as:
\begin{equation}\label{eq:singlefeature}
R_{j \to c} = 
\left| \frac{\partial y^{\, (L)}_c}{\partial x_j} \right|  
=
\left|\frac{\partial y^{\, (L)}_c}{\partial y_i^{\, (0)}}
\frac{\partial y_i^{\, (0)}}{\partial x_j} \right| \, .
\end{equation}

The partial derivative $\partial y_i^{\,(0)}/\partial x_j$ of Eq.~\eqref{eq:singlefeature} relates to the way $x_j$ is represented in the input. The process of creating an input from elements of $\mathcal{X}$ is what we term feature encoding. An input $y_i^{\,(0)}$ obtained from the feature $x_j$ is hence written as $y_i^{\,(0)} = g_i(x_j)$, where $g_i$ is the encoding function for the $i^{th}$ input. Considering the encoding process as such, we can write:
\begin{equation}
R_{j \to c} = 
\left|
 \frac{\partial y^{\, (L)}_c}{\partial y_i^{\, (0)}}
 \frac{\partial g_i(x_j)}{\partial x_j}
\right| \, .
\end{equation}

Notice how the importance depends on both the network, represented in the derivative from output to input, and the feature encoding process, given the presence of the encoding function.
The modulus operation ensures that the importance is always positive and real-valued. Here, we assume the network to be derivable in the vicinity of the current input, which might not be the case for some CVNN architectures.

Next, we examine the scenario where two arbitrary features, $x_j$ and $x_k$, are represented using a single input $y_i^{\, (0)}$, comparing their relevance. We define the relative importance between $x_j$ and $x_k$ to the $c^{\text{th}}$ output, $R_{j,k \to c}$ as the following ratio:
\begin{equation}\label{eq:relative_importance}
R_{j,k \to c} = 
\frac{ R_{j \to c} }{ R_{k \to c} } = 
 \left|\frac{\frac{\partial g_i(x_j,x_k)}{\partial x_j}}{\frac{\partial g_i(x_j,x_k)}{\partial x_k}}\right| \,.
\end{equation}

We see that the component of Eq.~\eqref{eq:singlefeature} related to the network is canceled, leaving only the derivatives of the encoding function. Thus, $R_{j,k \to c}$ is solely determined by the way features are encoded into $y_i$, and hence it is independent of the considered output. To simplify the notation, we drop the subscript indicating the output for the rest of this paper.
One of the consequences of Eq.~\eqref{eq:relative_importance} is that the encoding function chosen to combine $x_j$ and $x_k$ defines how these features are perceived by the model relative to one another. 

Although encoding functions are a method of pre-processing features, in the context of PNNs they can also be implemented in hardware. The incorporation of encoding functions in the circuit is particularly interesting for low-latency applications,  as the speed at which inputs are transformed and combined would be limited only by the reconfigurability of the driving electronics.
We now explore two types of complex encoding functions to see how they dictate relative feature importances. We also point out how they could be implemented in hardware.

\subsection{Exponential Encoding}

Since we are dealing with complex-valued inputs, one intuitive way to encode two features $x_j$ and $x_k$ into a single input would be to encode $x_j$ in its amplitude and $x_k$ in its phase. This encoding function can be written as:
\begin{equation}\label{eq:exponential_encoding}
    g(x_j, x_k) = x_je^{i x_k}  \, .
\end{equation}

The relative importance between these two features is calculated as:
\begin{equation}
R_{j,k \to c} =
\left|\frac{e^{ i x_k }}{i x_je^{i x_k}}\right| = 
\frac{1}{\left| x_j\right|} \, .
\end{equation}

In this case, the relative importance between the two features is dynamic, establishing an amplitude-dependent relation between the importances of amplitude and phase.

A hardware version of an exponential encoding function is shown in Fig.~\ref{fig:nsphere:big}(a), where a balanced MZI and a phase shifter are used to modulate the amplitude and phase of an input, respectively. 
The encodings and importances are not exactly the same since this amplitude modulation scheme is mediated by a sine function, it implements $g(x_j, x_k) = i \sin(x_j)\exp(x_k i)$. Here, Eq.~~\eqref{eq:exponential_encoding} can be achieved short of a global phase shift by mapping $x_j$ to $\arcsin(x_j)$.

\subsection{Linear Encoding}

Another way to combine two features is by encoding one in the real part and the other in the imaginary part of a complex input. This can be represented by the function:
\begin{equation} \label{eq:linearencoding}
    g(x_j, x_k) = x_j + i x_k \,.
\end{equation}

Here, we find their relative importance to be:
\begin{equation}
R_{j,k} = \frac{1}{\left|i\right|} = 1 \, ,
\end{equation}
thus indicating that both features will be considered to have the same importance for the network. Since they are independent from the weights of the network, their relative importance cannot be unlearned, i.e. it cannot be modified by further training. This might pose problems when the chosen encoding leads to relative importances that do not match the data.

An encoding function similar to that of Eq.~\eqref{eq:linearencoding} implemented in hardware, can be seen in Fig.~\ref{fig:nsphere:big}(b). There, two MZIs are used as amplitude modulators, while one of their outputs has its phase shifted by $\pi/2$ to encode the respective input in the imaginary axis. In that case, $g(x_j, x_k) = i(\sin(x_j) + \sin(x_k)i)$. Eq.~\eqref{eq:linearencoding} can be achieved short of a global phase shift by mapping $x_j$ and $x_k$ to $\arcsin(x_j)$ and $\arcsin(x_k)$.

\section{On the Impact of Encoding Functions to ANNs}
\label{n-sphere}    
    In this section, we address the practical implications of the discussions brought up in Section \ref{Methodology}.
Here, our objective is to demonstrate how a well-engineered encoding function can significantly improve the accuracy of an ANN on a test task. We begin by defining such task and studying the relative feature importances found in a solution to it. Later, we create an encoding function that reproduces these importances on trained ANNs, finally comparing its use against others.

Consider a simple classification problem with a known solution in the real domain: determining whether points lie inside or outside an $n$-sphere. 
An $n$-sphere is the generalization of a circle to $n+1$ dimensions, similar to how hyperplanes generalize planes. It is defined by a set of points $S^{(n)}$ that are equidistant from a central point $c_0 = (c_1, \cdots, c_{n+1})$ by a radius $r_0$. The distance of a point $P = (x_1, \cdots, x_{n+1})$ to $c_0$ is:
\begin{equation}\label{eq:n-sphere:n_sphere}
    r_p(x_1, ..., x_{n+1})  = \sqrt{\sum_{i=1}^{n+1} (x_i - c_i)^2}\,.
\end{equation}

Naturally, $P$ is considered outside of the $n$-sphere if $r_p$ exceeds $r_0$, and inside otherwise.
In this context, a mathematical model that outputs a probability of P being outside of $S^{(n)}$ can be constructed using a logistic function.
The logistic function $\sigma(x) = 1/(1 + e^{-x})$ is bounded between 0 and 1 with a smooth sigmoid transition, and is typically used in binary classification problems. Given the coordinates of $P$, this model can be expressed as:
\begin{equation}\label{eq:n-sphere:model}
    y = f (x_1, ..., x_{n+1}) = \sigma \left( r_p - r_0 \right)\,.
\end{equation}

Here, $y$ represents the probability that $r_p > r_0$ given the coordinates of $P$. When $y = 0.5$, Eq.~\eqref{eq:n-sphere:model} delineates the boundary defined by $S^{(n)}$, allowing for accurate classification of points based on this threshold. 

Since Eq.~\eqref{eq:n-sphere:model} can be used to accurately classify any point $P$, we conjecture that its relative feature importances are desirable to other models that wish to do it as well. Thus, we examine the sensitivity of $y$ to an arbitrary feature $x_j$, which can be calculated according to Eq.~\eqref{eq:singlefeature} as:
\begin{align}
\begin{split}
   R_j &= \left| \frac{\partial y}{\partial x_j} \right| \\
   &=  \left| \sigma(r_p - r_0)(1-\sigma (r_p - r_0)) \frac{x_j - c_j}{r_p} \right| \forall \, r_p\neq0  \,.
\end{split}
\end{align}

As can be seen in Fig.~\ref{fig:nsphere:big}(c), the importance of a feature peaks when it is at the $n$-sphere's boundary. Exactly at that point, small variations in $x_j$ cause the largest deviations of the probability of $P$ being outside of $S^{(\,n)}$. The relative feature importance between two features $x_j$ and $x_k$ is:
\begin{equation}\label{eq:nsphere:relative_golden}
   R_{j,k}  =  \left| \frac{x_j - c_j}{x_k - c_k} \right| \,.
\end{equation}

\begin{figure*}[t]
    \centering
    \includegraphics[width=\linewidth]{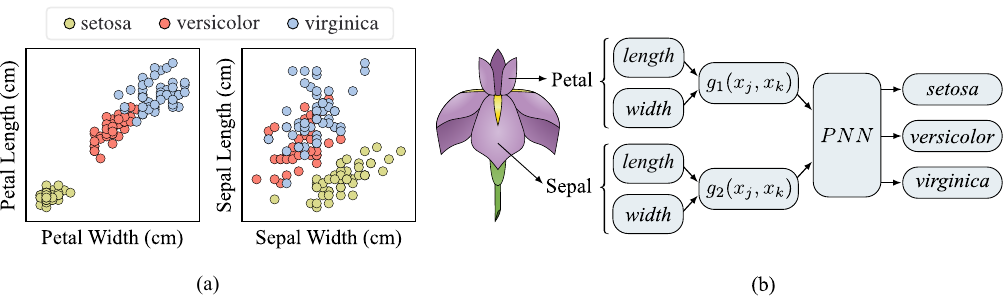}
    \caption{(a) Class distribution of normalized features of the Iris dataset (b) Schematic of the process of combining features utilized in the training of PNNs.}
    \label{fig:iris:scatter_schema}
\end{figure*}  
    
Now assume that, as in section \ref{Methodology}, we are constrained by size, and thus wish to combine different features to reduce the number of inputs. However, for this particular example (and by design), we have prior knowledge of the relative importances of features before combining them.
Taking advantage of this, we propose an encoding function $g(x_j, x_k)$ that achieves the desired reduction in dimentionality while preserving the relationships given by Eq.~\eqref{eq:nsphere:relative_golden}.
Given that after combining features, their relative importances should follow Eq.~\eqref{eq:relative_importance}, we can obtain one such $g(x_j,x_k)$ by solving the following system of partial derivatives:
\begin{equation}\label{eq:nsphere:partial_derivatives}
    \left\{\begin{matrix}
    \left| \frac{\partial g(x_j,x_k)}{\partial x_j}  \right| = \left| x_j - c_j  \right|\,, \\ \\
    \left| \frac{\partial g(x_j,x_k)}{\partial x_k}  \right|  = \left| x_k - c_k \right| \,.
\end{matrix}\right.
\end{equation}
This system can lead to one of many solutions of the form
\begin{equation}\label{eq:nsphere:real_representation}
    g(x_j,x_k) = \frac{1}{2}(x_j^2 + x_k^2) -c_j x_j - c_k x_k + C\,,
\end{equation}
where C is a constant.

In the same manner, we can obtain other functions that express different relative importances. A constant $R_{j,k} = 1$, for instance, is achieved by using $g(x_j, x_k) = (x_j + x_k)^n$ $\forall\, n \in \mathbb{R}$. Alternatively, $g(x_j, x_k) = (x_j \times x_k)^n $ $\forall\, n \in \mathbb{R}$ leads to  $R_{j,k} = |x_k/x_j|$, which is the inverse of Eq.~\eqref{eq:nsphere:relative_golden} when $c_j$ and $c_k$ are zero.
To compare the use of these encoding functions, we trained several ANNs, benchmarking them on networks that do not combining inputs (called ``independent" here). We are particularly interested in the performance of Eq.~\eqref{eq:nsphere:real_representation}, which we call ``engineered" encoding function. The description of the training procedures is as follows.

A dataset of $1000$ points in 4 dimensions was created, where each coordinate value was randomly chosen between $-2$ and $2$. Points were labeled as either inside or outside of a $3$-sphere $S^{\,(3)}$ of radius $1$, centered at the origin $c_0 = (0,0,0,0)$, according to their position. In order to obtain a balanced dataset, we generated the same amount of points inside and outside of the sphere. The networks trained to solve this task, were composed of an input layer containing either $2$ or $4$ neurons (depending on the combination of features or not), a hidden layer of $6$ neurons and an output layer with a single neuron. A logistic activation function $\sigma$ was used for every layer. Each encoding function was used to train $100$ different networks, thus accounting for the random initialization of weights and random shuffling of the dataset prior to training. The networks were trained on $70\%$ of the available data for $100$ epochs with a learning rate of $0.001$, and tested on the remaining data. 

The results of these experiments are shown in Fig.~\ref{fig:nsphere:big}(d). We notice that some representations can render the task harder to solve, while others maintain, to some extent, the accuracy achieved by the use of independent inputs. The engineered encoding function in Eq.~\eqref{eq:nsphere:real_representation} outperformed all others.
With this example, we show that the way we combine features plays a role in the accuracy of ANNs. Given prior knowledge on how features relate to one another, which may come from domain-specific knowledge or from inspecting the data (noticing symmetries or class distributions, for example), we could estimate relative feature importances and obtain an encoding function that aligns with them. Combining features with said encoding function could improve the network performance.

\section{Application of Encoding Functions in PNNs}
\label{Simulations}
    In this section, we retake the subject of this study and explore the use of different encoding functions in PNNs trained on the Iris dataset \cite{fisher1936use}, a standard benchmark for classification algorithms. Our goal is to show how carefully chosen encoding functions might lead to higher accuracies in PNNs. To this end, we compare the performance of several encoding functions by means of simulations of PNNs, which differ significantly from the ANNs of the previous section in terms of their complex-valued inputs and transformations.

The Iris flower classification task involves categorizing three different Iris species (Setosa, Versicolour, and Virginica) based on four features: the lengths and widths of sepals and petals. The dataset, consisting of $150$ labeled data points, has considerable class overlaps, such that no single feature alone can distinguish all species, making this an ideal candidate for our experiments. Visualizations of feature distributions and class overlaps are presented in Fig.~\ref{fig:iris:scatter_schema}(a).

\begin{figure*}[t]
    \centering
    \includegraphics[width=\linewidth]{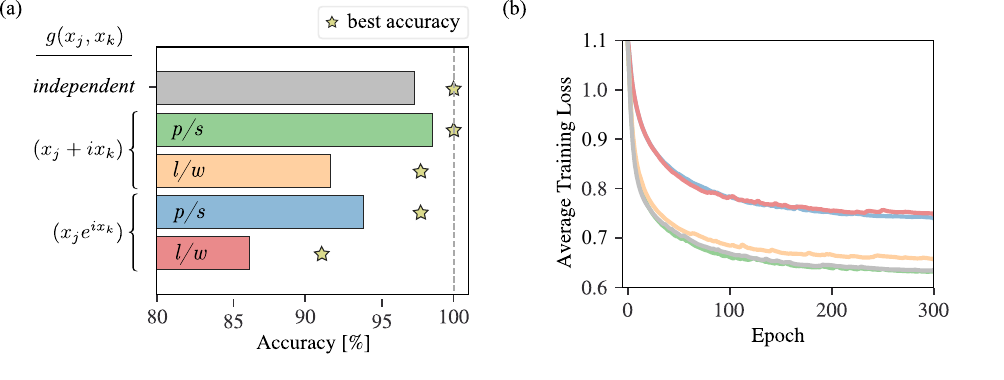}
    \caption{(a) Accuracy of PNNs for different encoding functions. ``independent" represents the accuracy of a PNN trained without combining features. (b) Average training loss of the trained PNNs per epoch. The color scheme is the same as the one defined in (a).}
    \label{fig:iris:accuracy_loss}
\end{figure*}  

Our experimental design involves training PNNs by combining features in pairs as illustrated in Fig.~\ref{fig:iris:scatter_schema}(b). We assess their performance by averaging the accuracy of $100$ trained PNNs, benchmarking them against a PNN that does not combine features. 
This sample size was chosen to allow for convergence in the average values obtained for accuracy, given the variability in the training process.
The architecture of the PNNs consists of a single hidden layer with $6$ neurons. Depending on the configuration, the number of input neurons varies between $3$ (when combining features) and $5$ (when using independent inputs), where one input acts as a bias for both configurations. The output layer has $3$ neurons, matching the number of classes.
All configurations use the same underlying circuit, where NN layers are implemented using meshes of MZIs with trainable phase shifters, as represented in Fig~\ref{fig:layerANNandPNN}(b).
Every layer is followed by a \textit{softplus} activation function, which can be implemented in integrated photonic circuits \cite{campo2022reconfigurable}. Although its hardware implementation would change both the modulus and the phase of the signals, we model it by applying $softplus(x) = \log(1+\exp(x))$ solely to the modulus of the complex numbers \cite{Banerjee2023}. This approach allows us to simulate the gain and activation behavior while simplifying the model by avoiding additional phase changes. These phase changes can make the simulation and training more challenging and are less critical to the primary function of the \textit{softplus} activation in this context.

The circuits were simulated using the Photontorch Python package \cite{laporte2019highly}. The simulations were performed under ideal conditions, excluding noise and component imperfections. They were trained for $300$ epochs on $70\%$ of the dataset, reserving the remaining $30\%$ for testing. The dataset was divided into five shuffled batches per epoch to enhance training stability. A \textit{softmax} function was used to convert the output light intensity values into class probabilities \cite{goodfellow2016deep}. Weight updates were performed using a cross-entropy loss function combined with stochastic gradient descent. The initial learning rate was set at $0.01$ and adjusted at learning plateaus. While higher accuracies may be achieved by further optimizing the training process to each specific case, we opted for a constant training procedure across all circuits to isolate the effects of different encoding functions.

Here, we investigate the use of the encoding functions detailed in Section \ref{Photonic Neural Networks}: linear and exponential encoding. Given the anisotropic nature of the Iris classification task, unlike the $n$-sphere problem, we also consider which features to combine. To explore the impacts of this choice on the obtained accuracy, we use two combination strategies for features: grouping by the lengths and widths $(l/w)$ or by petal and sepal information $(p/s)$.
We benchmarked the performance of PNNs using different encoding functions and groupings of features to the independent case, where features were not combined. 

The results of these experiments, shown in Fig.~\ref{fig:iris:accuracy_loss}, are summarized as follows.
Exponential encoding exhibited the lowest performance, falling up to $11\%$ in mean accuracy compared to the independent benchmark. In contrast, linear encoding, commonly used in the photonics community \cite{Banerjee2023, hamerly2022asymptotically, wang2022Multicore,Fang2019, qiu2024oplixnet}, was able to match the performance of the independent case. The difference between the best and worst performing encoding functions was $12.3\%$. These results highlight that both the manner in which features are combined and the combination of features itself play significant roles in the final accuracy of PNNs.
When comparing different feature groupings, we found that $(l/w)$ consistently performed worse than $(p/s)$, demonstrating that the choice of which features to combine can also impact accuracy for some tasks.

These findings are supported by heuristics found in the data. A closer inspection of Fig.~\ref{fig:iris:scatter_schema}(a) reveals that petal length and petal width together are highly discriminative of the different classes. These features separate different species in a similar fashion, as evidenced by the distribution of classes along the diagonal of the plot, suggesting that they may have similar importance. Thus, the combination ($p/s$) with linear encoding would combine petal length and width with an equal relative importance, expressing such relationships.

\section{Conclusions}
\label{Conclusions}
    Combining features into single inputs in PNNs can lead to reduced number of inputs and associated devices as well as enabling the use of smaller and more energy efficient NNs. These benefits would help to render some circuits more feasible to be simulated, fabricated, tested or deployed.
However, this method of feature combination imposes predefined relationships among the features that may not necessarily reflect the nature of data or task at hand.
Nonetheless, selecting or designing encoding functions based on an understanding of the dataset or from domain-specific knowledge can lead to improved accuracy. 
We have illustrated this first on an ideal simple example, and then for simulated PNNs.

In the scenarios shown here, as it is seen in literature, features are combined into a single input. As an alternative, we could distribute features across many inputs, circumventing the discussions brought up here and making it possible to learn other relative feature importances. 
For instance, Principal Component Analysis (PCA) can be used for dimensionality reduction, distributing features across many inputs simultaneously \cite{mojaver2023addressing}.
Expanding on this concept, a learnable encoding function that uses every feature available would be a fully connected layer of a NN \cite{wang2022Multicore}, which is more complex and less efficient than what is explored in our work. Besides, the approach used here could be applied directly at a hardware level, using integrated photonics and CMOS-compatible platforms for volume production.

Here, the discussions highlight that there is no neutral way of using this feature combination strategy in PNNs. Combining features in this manner will necessarily emphasize certain feature relationships. Sometimes, a PNN might achieve good performance metrics despite of combinations that are not ideal. However, even if high accuracy is achieved, these combinations can also introduce or amplify biases in the model outputs, depending on the specific features and their encoded interactions.
Rather than leaving this to chance, we suggest to carefully assess how to encode features given the nature of the problem and data.

\begin{acknowledgments}

The authors would like to acknowledge and thank \textbf{Peter Bienstman} and \textbf{Thomas Van Vaerenbergh} for their valuable advice and discussions during the early stages of this work, as well as for reviewing the paper before submission.

This project received funding from École Centrale de Lyon and ANR (No. ANR-20-THIA-0007-01). \textbf{Paul Jimenez} and \textbf{Fabio Pavanello} thank ANR's support (No. ANR-20-CE39-0004), and \textbf{Fabio Pavanello} acknowledges support by the European Union’s Horizon Europe research and innovation program (No.101070238).

Views and opinions expressed are however those of the author(s) only and do not necessarily reflect those of the European Union. Neither the European Union nor the granting authority can be held responsible for them.

\end{acknowledgments}
\section*{Author Declarations}

    \subsection*{Conflict of Interest Statement}
    
    The authors have no conflicts to disclose.
    
    \subsection*{Author Contributions}
    
    \textbf{Mauricio Gomes de Queiroz:} Conceptualization (lead); Data curation (lead); Formal analysis (lead); Investigation (lead); Methodology (lead); Software (lead); Visualization (lead); Writing/original draft preparation (lead); Writing/review \& editing (equal).
    \textbf{Paul Jimenez:} Formal analysis (supporting); Methodology (supporting); Writing/review \& editing (equal).
    \textbf{Raphael Cardoso:} Methodology (supporting); Writing/review \& editing (equal).
    \textbf{Mateus Vidaletti da Costa:} Methodology (supporting); Writing/review \& editing (equal).
    \textbf{Mohab Abdalla:} Methodology (supporting); Writing/review \& editing (equal).
    \textbf{Ian O'Connor:} Supervision (supporting); Writing/review \& editing (equal).
    \textbf{Alberto Bosio:} Funding Acquisition (equal); Supervision (supporting); Writing/review \& editing (equal).
    \textbf{Fabio Pavanello:} Conceptualization (supporting); Formal analysis (supporting); Methodology (supporting); Funding Acquisition (equal); Supervision (lead); Writing/review \& editing (equal).
    
    \subsection*{Data Availability Statement}
    
    The code that reproduces the experiments, and the data that support the findings of this study are available at \href{https://github.com/mgomesq/feature_representation_pnns}{github.com/mgomesq/feature\_representation\_pnns}.
    
\bibliography{references}

\end{document}